\pdfoutput=1

\documentclass[aps,prb,twocolumn,superscriptaddress,showpacs]{revtex4}

\usepackage{graphicx}

\begin{document}



\title{Voltage-controlled Group Velocity of Edge Magnetoplasmon\\
in the Quantum Hall Regime}



\author{H. Kamata}
\affiliation{NTT Basic Research Laboratories, NTT Corporation, 
3-1 Morinosato-Wakamiya, Atsugi, Kanagawa 243-0198, Japan}
\affiliation{Research Center for Low-Temperature Physics, Tokyo 
Institute of Technology,
2-12-1 Ookayama, Meguro, Tokyo 152-8551, Japan}
\author{T. Ota}
\author{K. Muraki}
\affiliation{NTT Basic Research Laboratories, NTT Corporation, 
3-1 Morinosato-Wakamiya, Atsugi, Kanagawa 243-0198, Japan}
\author{T. Fujisawa}
\affiliation{Research Center for Low-Temperature Physics, Tokyo 
Institute of Technology,
2-12-1 Ookayama, Meguro, Tokyo 152-8551, Japan}


\date{\today}

\begin{abstract}
We investigate the group velocity of edge magnetoplasmons (EMPs) 
in the quantum Hall regime by means of time-of-flight measurement.
The EMPs are injected from an Ohmic contact by applying a voltage pulse, 
and detected at a quantum point contact by applying 
another voltage pulse to its gate.
We find that the group velocity of the EMPs traveling along the edge 
channel defined by a metallic gate electrode strongly depends 
on the voltage applied to the gate.
The observed variation of the velocity can be understood to reflect 
the degree of screening caused by the metallic gate,
which damps the in-plane electric field and hence reduces the velocity.
The degree of screening can be controlled by changing the distance between 
the gate and the edge channel with the gate voltage.
\end{abstract}


\maketitle


\section{Introduction}

When a strong magnetic field is applied perpendicular to the two-dimensional 
electron gas (2DEG), electrons propagate in one-dimensional edge channels, 
which form along the edge of the sample in the quantum Hall 
regime. \cite{ezawa:QHE}
Recently, edge channels have attracted much attention as a coherent 
one-dimentional channel without dissipation.
For example, electronic Hanbury Brown-Twiss, \cite{henny:1999} 
Mach-Zehnder, \cite{yang:2003} and Fabry-P\'{e}rot \cite{yiming:2009, 
McClure:2009, Ofek:2009} interferometers have been realized experimentally 
by employing edge channels together with quantum point contacts (QPCs) 
to split and join them.
These interferometric experiments allow us to study coherent transport and 
quantum statistics of electrons. \cite{neder:2007-1, neder:2007-2, 
roulleau:2008}
On the other hand, a single-electron source has been realized by tailoring 
the time-dependent electrochemical potential
of a quantum dot connected to an edge channel. \cite{feve:2007}
These experiments motivate the study of electronic quantum channels, which 
carry quantum states over a long distance extending over the whole device.
For such purposes, the electron velocity is one of the important 
characteristics to control the timing of the transmission. \cite{feve:2008}

When non-equilibrium excess charge is induced in the edge channel, for 
example, by applying a voltage pulse or microwave, the charge propagates 
along the edge of the sample in the form of an edge magnetoplasmon (EMP).
EMPs have been widely investigated both theoretically \cite{fetter:1985, 
volkov:1988, aleiner:1994, johnson:2003} 
and experimentally. \cite{ashoori:1992, ernst:1996, sukhodub:2004}
For example, high-frequency magnetoconductivity measurements and 
time-of-flight experiments have shown that the group velocity of the EMPs 
is inversely proportional to the magnetic field $B$. \cite{ashoori:1992, 
ernst:1996, sukhodub:2004}
The reported velocity of the EMPs is $1000 \sim 1500$ km/s at bulk filling 
factor $\nu = 2$ when the edge channel is formed along a chemically etched 
mesa structure. \cite{ashoori:1992, ernst:1996}
On the other hand, a much smaller velocity ($\sim 150$ km/s at $\nu = 2$) 
was observed when the surface of the structure was covered by
a metal. \cite{sukhodub:2004}
Although the difference is believed to be due to image charge or screening 
by the surface metal, \cite{johnson:2003}
no systematic study has been carried out in the intermediate region.

In this paper, we investigate the group velocity of EMPs traveling along 
the edge channel defined by a semi-infinite metallic gate.
We find that the group velocity strongly depends on the gate voltage ($280 
\sim 430$ km/s at $\nu = 2$).
The observed gate voltage dependence can be qualitatively understood by 
considering the degree of screening.
The electrically tunable group velocity is attractive for controlling 
the timing of EMP transport.

\section{Time-of-flight measurement with a quantum point contact}

\begin{figure}[h]
\begin{center}
\includegraphics[width=71mm]{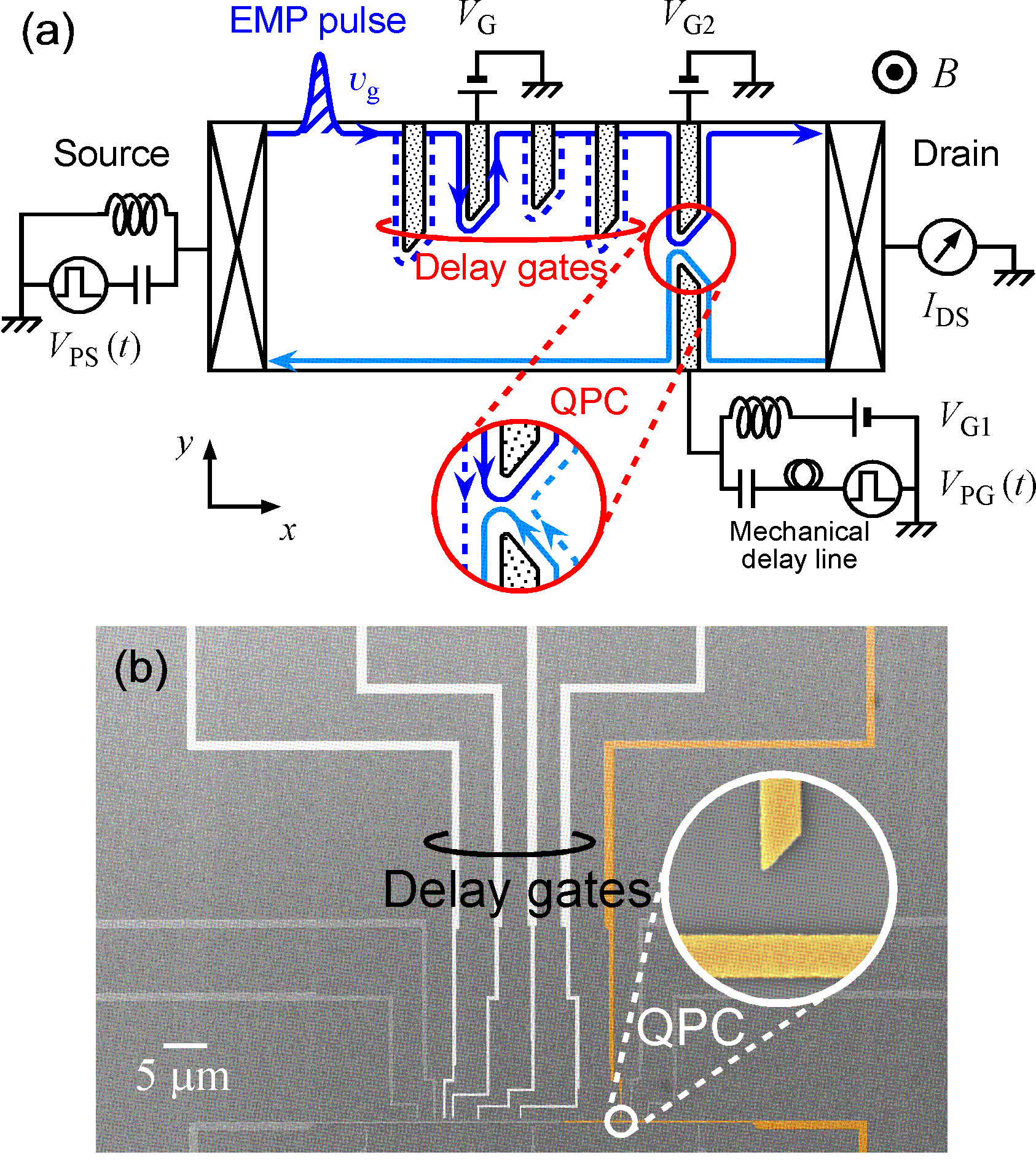}
\caption{(Color online) (a) Schematic device structure and experimental
setup for time-of-flight measurement.
A short voltage pulse $V_{\mathrm{PS}} (t)$ of 1.0 mV in amplitude is 
applied to the source to inject a pulse of EMPs.
Another voltage pulse $V_{\mathrm{PG}} (t)$ of 20 mV in amplitude is 
applied to the QPC to probe the local potential.
The time interval between the two voltage pulses is changed by the 
mechanical delay line.
Four delay gates between the source and the QPC can be used to add 
extra path length.
(b) A scanning electron micrograph of the device.
The orange and white lines are metallic gates for the QPC and the delay 
gates, respectively.
The light gray lines are unused.
Disabled gates are biased at $\sim + 0.2$ V to minimize backscattering.
The complicated gate patterns are designed for another purpose, but their 
different perimeters are useful for this work.}
\label{fig_setup}
\end{center}
\end{figure}

In previous measurements of EMPs in the frequency or time domains,
EMPs were detected with relatively large Ohmic contacts 
\cite{ernst:1996, sukhodub:2004} or gate electrodes. \cite{ashoori:1992}
Here, we use a QPC as a local probe of the charge or the associated
potential created by EMPs.
This experimental technique was originally developed for evaluating 
time-dependent potentials induced by external voltage 
pulses. \cite{kamata:2009}
In this paper, we utilize this technique to study charge dynamics of
EMPs.

Figure \ref{fig_setup}(a) schematically shows the device structure and 
experimental setup.
The structure, fabricated on an AlGaAs/GaAs modulation-doped 
heterostructure, consists of a QPC defined by a standard split-gate 
technique and four additional metallic gates serving as
``delay gates" to tune the path length (used in Sec.~III).
The 2DEG located 110 nm below the surface has a density of 
$3.2 \times 10^{15}$ m$^{-2}$ and a low-temperature mobility of 
$170$ m$^{2}$/Vs.
The following measurements were performed in a dilution refrigerator 
at about 50 mK.
A constant magnetic field of $B = 6.5$ T was applied to the 2DEG, which 
corresponds to a bulk filling factor $\nu = 2$.

\begin{figure}[h]
\begin{center}
\includegraphics[width=77mm]{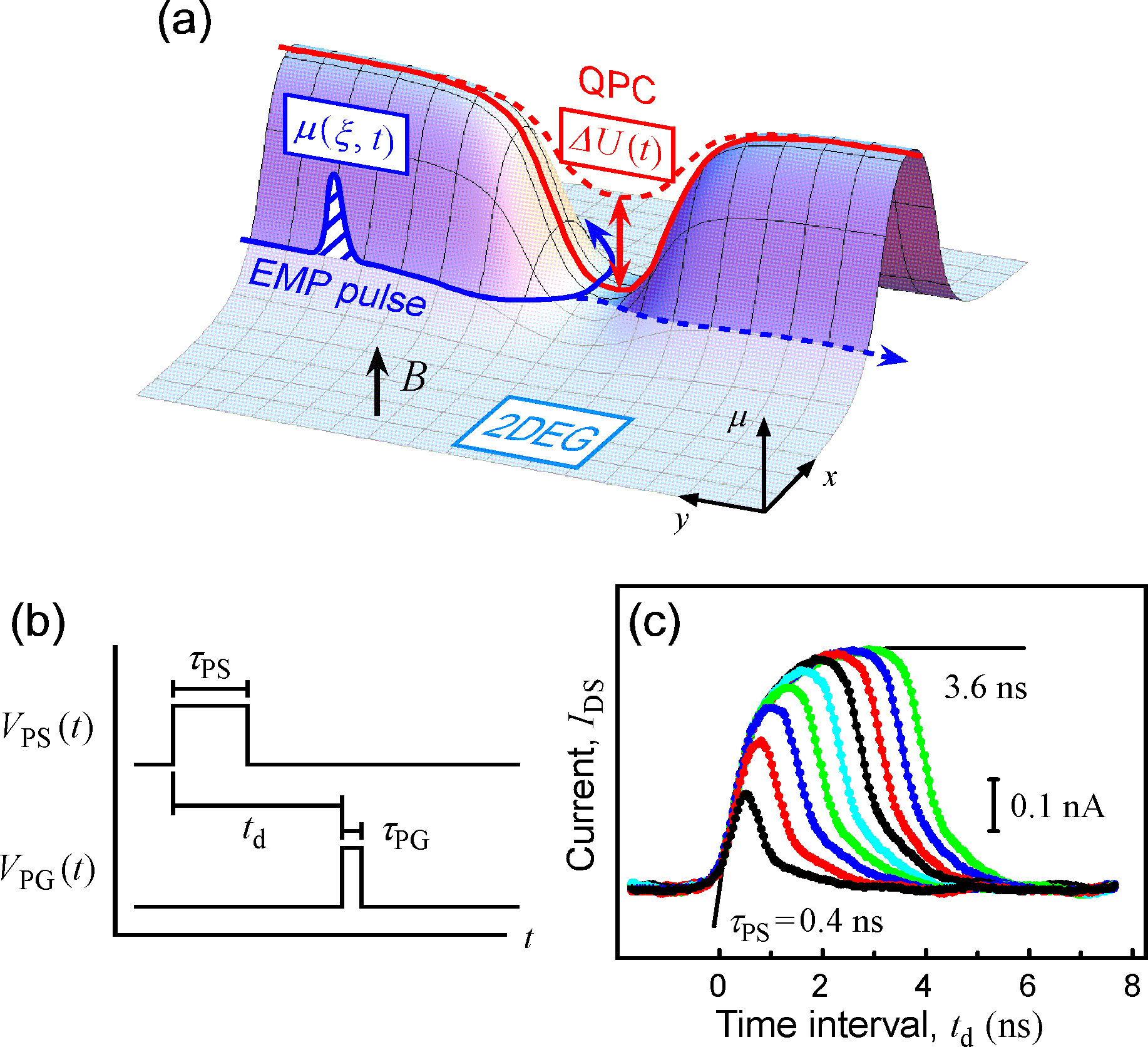}
\caption{(Color online) (a) Schematic potential diagram of the EMP
pulse and the QPC.
If the timing of the voltage pulse applied to the QPC coincides 
with the arrival of the EMP pulse at the QPC, the EMP pulse passes
through the QPC (solid line); otherwise, the EMP pulse is 
reflected at the QPC and returns to the source contact (dashed line).
(b) Schematic pulse patterns of the two voltage pulses.
(c) $I_{\mathrm{DS}} (t_{\mathrm{d}})$ curves observed at $B = 6.5$ T
for various pulse widths.
The solid line is  an exponetial fit.}
\label{fig_QPC}
\end{center}
\end{figure}

When a voltage pulse, $V_{\mathrm{PS}} (t)$, is applied to the
source Ohmic contact, a pulse of EMPs is generated and then
travels chirally along the edge channel as schematically shown 
in Fig.~\ref{fig_setup}(a).
The charge in the form of the EMP pulse occupies the Landau Levels
up to the electrochemical potential $\mu(\xi$, $t)$, which depends 
on the distance ($\xi$) along the edge channel from the source.
Fig.~\ref{fig_QPC}(a) schematically shows the potential 
$\mu(\xi$, $t)$ around the QPC for a single Landau level, with
the formation of a compressible strip neglected.
If the dispersion of the EMP pulse is approximated to be a linear 
function with a constant velocity $v$, the dynamical potential 
can be written in the simple form $\mu(\xi$, $t) = 
\mu^{\prime} (t - \xi / v)$.
Then, the potential arrives at the QPC (at a distance 
$\xi_{\mathrm{Q}}$) with a delay time of $\xi_{\mathrm{Q}} / v$ 
after the application of the voltage pulse to the source.
Another voltage pulse, $V_{\mathrm{PG}} (t - t_{\mathrm{d}})$, 
is applied to the lower gate of the QPC in the tunneling regime to 
probe the local potential.
Here, $t_{\mathrm{d}}$ is the time interval between the two voltage 
pulses, which can be experimentally controlled with a mechanical 
delay line and a pulse pattern generator as shown in 
Fig.~\ref{fig_QPC}(b).
The pulse applied to the QPC changes the barrier potential of the 
QPC, $U (t)$, and hence the conductance, $G_{\mathrm{Q}} (t)$, 
by $\mathit{\Delta} G_{\mathrm{Q}} (t) \propto \mathit{\Delta} U (t) 
\propto V_{\mathrm{PG}} (t)$ for sufficiently small amplitude of 
the pulse.
In the linear transport regime, the current through the QPC, $i (t)$,
is proportional to the potential difference $\mathit{\Delta} \mu (t)
= \mu (\xi_{\mathrm{Q}}$, $t) - \mu_{\mathrm{D}}$ across
the QPC, where $\mu_{\mathrm{D}}$ is the (time-independent) 
electrochemical potential of the drain.
Instead of measuring $i (t)$ directly, we measured the average 
current $I_{\mathrm{DS}} (t_{\mathrm{d}}) = \langle i (t) \rangle$ 
as a function of $t_{\mathrm{d}}$.
Then, the average current has a form of correlation function,
\begin{equation}
I_{\mathrm{DS}} (t_{\mathrm{d}})
= \frac{1}{e} \langle
\mathit{\Delta} G_{\mathrm{Q}} (t - t_{\mathrm{d}})
\mathit{\Delta} \mu (t)
\rangle.
\label{CorrelationFunction}
\end{equation}
One can evaluate the time-dependent local potential
$\mathit{\Delta} \mu (t)$ if $\mathit{\Delta} G_{\mathrm{Q}} (t)$
abruptly changes in a delta-function $\mathit{\Delta} 
G_{\mathrm{Q}} (t) \propto \delta (t)$ or in any known functions.
The actual potential waveforms in the device can be
estimated by analyzing $I_{\mathrm{DS}} (t_{\mathrm{d}})$
for various pulse widths. \cite{kamata:2009}

The $I_{\mathrm{DS}} (t_{\mathrm{d}})$ curves shown in 
Fig~\ref{fig_QPC}(c) were obtained by varying the pulse width 
$\tau_{\mathrm{PS}}$ of $V_{\mathrm{PS}} (t)$ from 0.4 to 
3.6 ns while keeping the pulse width $\tau_{\mathrm{PG}}$
of $V_{\mathrm{PG}} (t)$ constant at 0.08 ns
(minimum available pulse width in our setup).
The peak width in the observed $I_{\mathrm{DS}} (t_{\mathrm{d}})$ 
curve increases with pulse width $\tau_{\mathrm{PS}}$.
Since $\mathit{\Delta} G_{\mathrm{Q}} (t)$ effectively changes
in a delta-function,
the observed $I_{\mathrm{DS}} (t_{\mathrm{d}})$ curves reflect
the time evolution of the EMP pulse in the device.
Here, the time constant of 0.6 ns, which is obtained by 
fitting with an exponetial function [Fig~\ref{fig_QPC}(c), 
solid line] is larger than that measured at zero magnetic field 
(0.28 ns). \cite{kamata:2009}
The increased time constant may be related to the higher 
Ohmic resistance in the magnetic field or to the non-linear 
dispersion of EMPs, which will be investigated 
in the future.
In this way, we can evaluate the time-dependent potential or 
the charge distribution of the EMP pulse.
In this paper, we focus on the velocity of the EMPs.
The following data were measured at $\tau_{\mathrm{PS}} = 0.4$ ns
and $\tau_{\mathrm{PG}} = 0.08$ ns.

\section{Experimental Results}

Edge channels can be defined either by chemically etching the
heterostructure or electrostatically depleting the 2DEG under 
a metallic gate.
Edge channels defined by a metallic gate are useful for electrical 
switching of the path length.
Here, we use four additional gates (delay gates; 
typically $\sim 100$ $\mu$m in length) between the source contact 
and the QPC [Fig.~\ref{fig_setup}(b)], which add extra path length 
(the perimeter), $\mathit{\Delta} L$, by depleting electrons 
underneath with a negative gate voltage $V_{\mathrm{G}}$ 
(below the pinch-off voltage of $\sim - 0.2$ V).
Since these gates are slightly different in perimeter, we can choose 
16 kinds of path lengths by combining them.

\begin{figure}[h]
\begin{center}
\includegraphics[width=74mm]{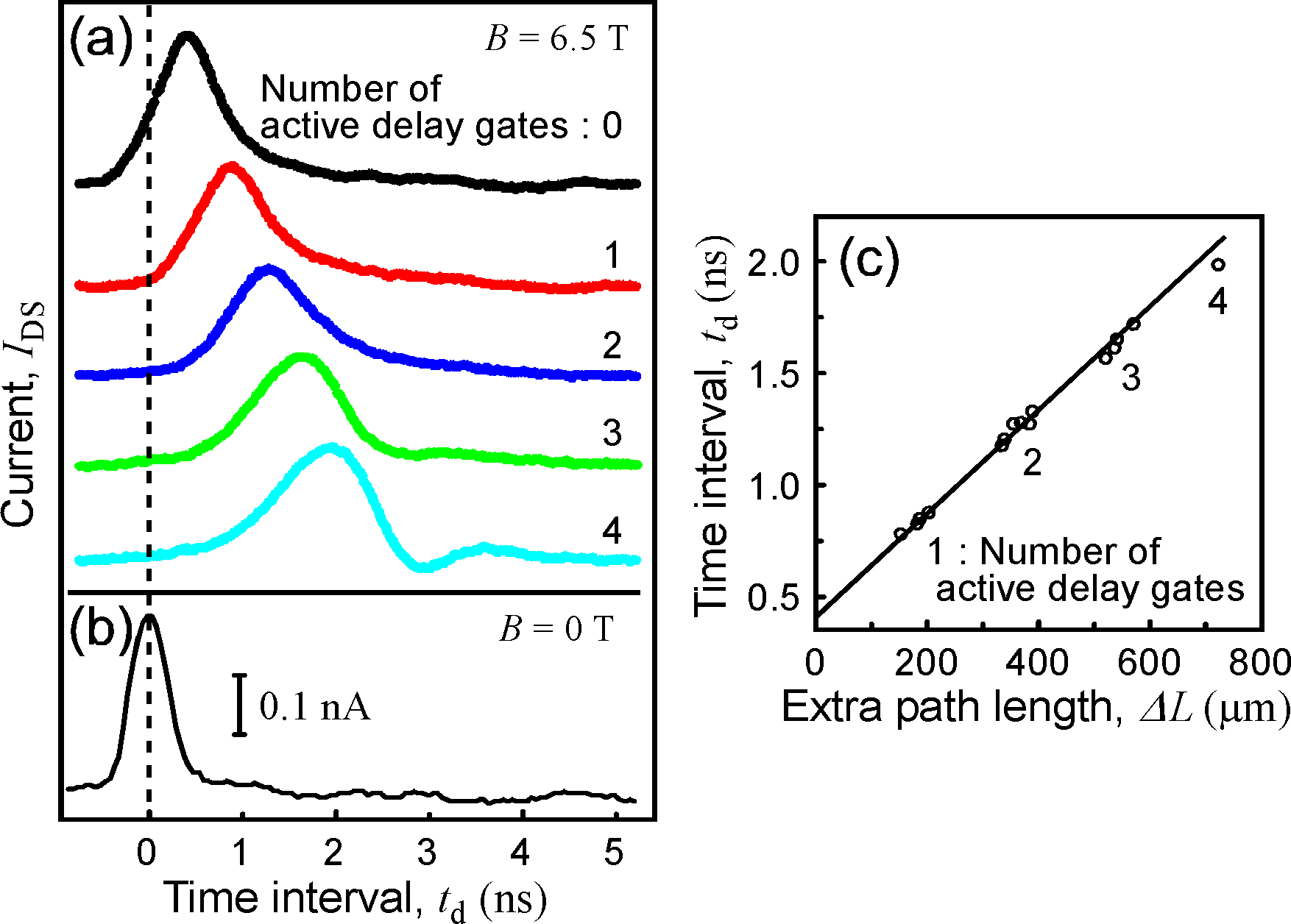}
\caption{(Color online) (a) and (b) $I_{\mathrm{DS}} 
(t_{\mathrm{d}})$ curves 
observed at (a) $B = 6.5$ T ($\nu = 2$) and (b) $B = 0$ T.
The origin of the time interval is chosen at the peak position of 
the reference data in (b).
Active delay gates are biased at $V_{\mathrm{G}} = - 1.1$ V.
These curves are offset for clarity.
(c) Time interval for various extra path length $\mathit{\Delta} L$ 
with a straight-line fit.}
\label{fig_length}
\end{center}
\end{figure}

The top curve in Fig.~\ref{fig_length}(a) shows the 
$I_{\mathrm{DS}} (t_{\mathrm{d}})$ observed 
at $B = 6.5$ T ($\nu = 2$) when no delay gates are activated 
(the number of the active delay gate is 0). 
As compared with the reference data observed at zero magnetic 
field [Fig.~\ref{fig_length}(b)], 
the cureves observed at $B = 6.5$ T [Fig.~\ref{fig_length}(a)]
is significantly delayed.
When no delay gates are activated, the EMP pulse injected from 
the source travels along the chemically etched edge 
($\sim 400$ $\mu$m in length) and the edge defined by 
a metallic gate (the upper gate of the QPC; $\sim 100$ $\mu$m in 
length) until it reaches the QPC detector.
The observed delay time of $t_{\mathrm{d}} \sim 0.5$ ns corresponds 
to the mean group velocity of $\sim 1000$ km/s, which is consistent
with the previous reports. \cite{ashoori:1992, ernst:1996}

As the number of active delay gates ($V_{\mathrm{G}} = - 1.1$ V) and 
hence the extra path length $\mathit{\Delta} L$ is increased, the 
delay of the EMP pulse increases as shown in Fig.~\ref{fig_length}(a). 
The delay time $t_{\mathrm{d}}$ is proportional to 
$\mathit{\Delta} L$ as 
shown in Fig.~\ref{fig_length}(c).
From the slope, the group velocity of the EMP pulse traveling along 
the delay gates can be precisely determined to be $v_{\mathrm{g}} 
= 430$ km/s at $V_{\mathrm{G}} = - 1.1$ V.
The obtained velocity is smaller than the reported value in the 
unscreened case \cite{ashoori:1992, ernst:1996} but 
larger than that in the high-screening case. \cite{sukhodub:2004}

\begin{figure}[h]
\begin{center}
\includegraphics[width=60mm]{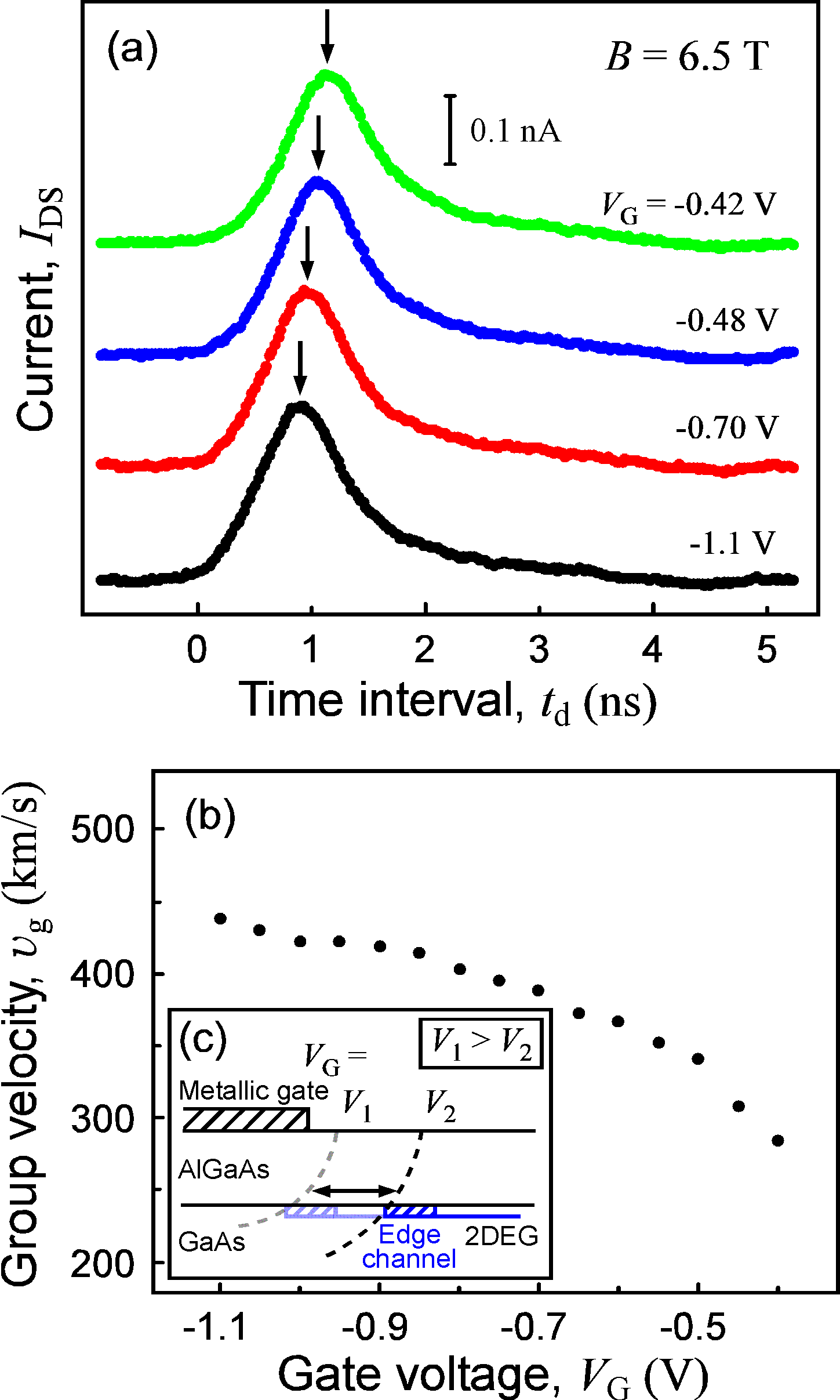}
\caption{(Color online) (a) $I_{\mathrm{DS}} (t_{\mathrm{d}})$ 
curves observed for various $V_{\mathrm{G}}$.
These curves are offset for clarity.
(b) Group velocity $v_{\mathrm{g}}$ as a function of 
$V_{\mathrm{G}}$.
(c) Schematic cross-section of the heterostructure.
The location of the edge channel is shifted away from the gate 
as the 2DEG is depleted.}
\label{fig_VG}
\end{center}
\end{figure}

We find that the velocity strongly depends on the gate 
voltage $V_{\mathrm{G}}$.
Fig.~\ref{fig_VG}(a) shows the variation of 
$I_{\mathrm{DS}} (t_{\mathrm{d}})$ observed when only one of 
the delay gates ($\mathit{\Delta} L = 205$ $\mu$m)
is activated at different $V_{\mathrm{G}}$.
The peak position indicated by the arrows shifts with 
$V_{\mathrm{G}}$.
By measuring the $\mathit{\Delta} L$-dependence of 
$t_{\mathrm{d}}$ 
at various $V_{\mathrm{G}}$, we obtain the velocity as
a function of $V_{\mathrm{G}}$ as shown in Fig.~\ref{fig_VG}(b).
As $V_{\mathrm{G}}$ is made more negative, the depletion region 
spreads as illustrated in Fig.~\ref{fig_VG}(c) and hence
the edge channel is pushed further away from the metallic gate.
Therefore, the observed variation of $v_{\mathrm{g}}$ is 
considered to reflect the degree of screening by the metallic
gate.

Here, several points need to be addressed.
First, we note that the path length slightly changes with 
$V_{\mathrm{G}}$ since the distance between the edge channel
and the gate changes with $V_{\mathrm{G}}$.
However, the change (estimated to be a few 100 nm) is negligibly 
small as compared with the total path 
length (a few 100 $\mu$m).
Moreover, the path length increases as $V_{\mathrm{G}}$ is made 
more negative, which does not account for the earlier arrival of
the EMP pulse.
Second, the delay gates have complicated shapes with varying 
widths ($0.1 \sim 1$ $\mu$m) and corners as shown in 
Fig.~\ref{fig_setup}(b).
Although this may cause the screening strength to vary locally,
the observed linearity to $\mathit{\Delta} L$ 
[Fig.~\ref{fig_length}(c)] and 
smooth variation with $V_{\mathrm{G}}$ 
[Fig.~\ref{fig_VG}(b)] ensure that this effect is minor.
Third, electrostatics shows that, at the bulk filling factor 
$\nu = 2$, the edge of the 2DEG consists of two wide compressible 
strips separated by, a much narrower, incompressible strip with 
local filling factor $\nu = 1$. \cite{chklovskii:1992}
The data shown in this paper, which were taken at $0 < 
G_{\mathrm{Q}} < e^{2} / h$, should in principle correspond to 
the charge distribution in the outer compressible strip.
However, no qualitative difference was observed when 
$G_{\mathrm{Q}}$ was set at $e^{2} / h < G_{\mathrm{Q}} 
< 2 e^{2} / h$ to detect the charge distribution in the inner 
compressible strip.
This suggests that the EMP pulse is spread over the two compressible 
strips and its properties may be well characterized by the edge 
channel of the lowest spin-unresolved Landau level. 
\cite{sukhodub:2004}

\section{Discussions}

\begin{figure}[h]
\begin{center}
\includegraphics[width=65mm]{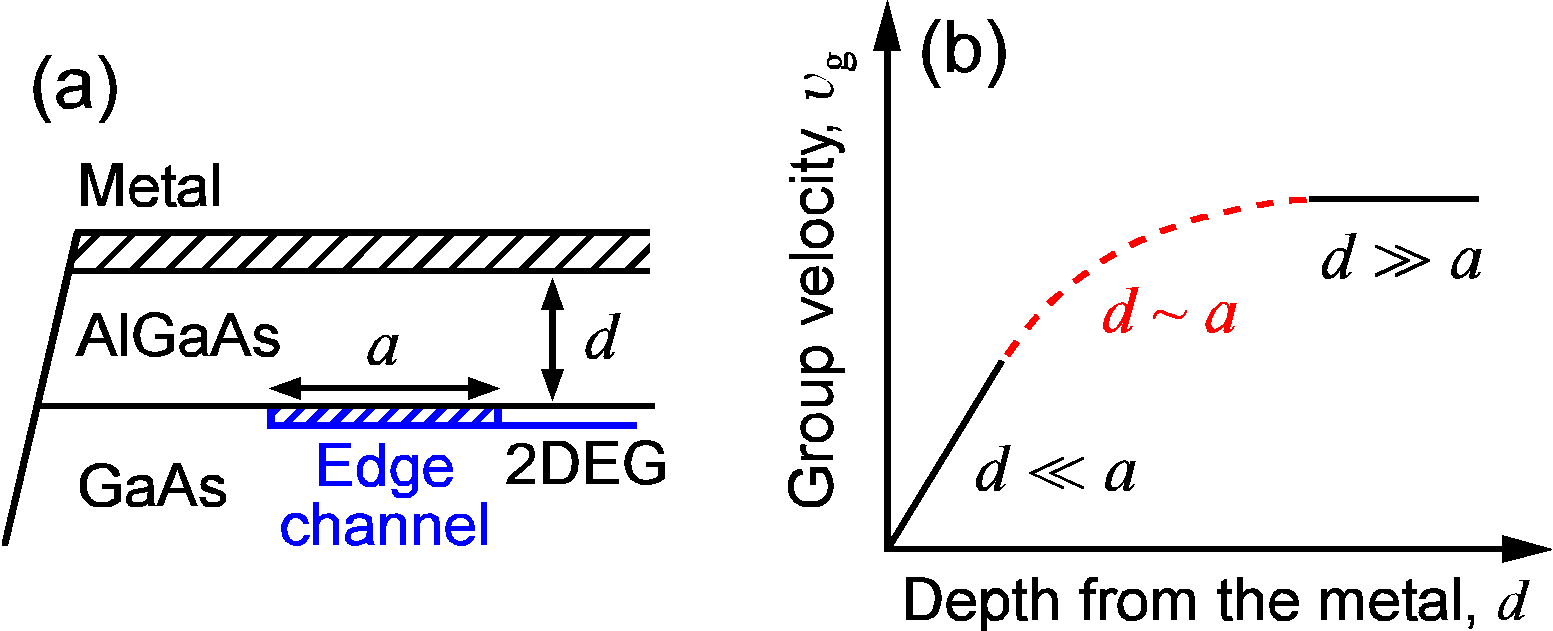}
\caption{(Color online) (a) Schematic cross-section of a 
heterostructure covered with a metal.
(b) Schematic illustration of the group velocity depending on
the depth.}
\label{fig_screening}
\end{center}
\end{figure}

The dispersion relation and the group velocity $v_{\mathrm{g}}$ of 
EMPs have been investigated theoretically in the
long-wavelength limit for different boundary conditions.
In the case of unscreened 2DEG for the etched edge, 
\cite{aleiner:1994} the velocity $v_{\mathrm{g}}$ at wave 
number $k$ for 
the fundamental EMP mode is given by
\begin{equation}
v_{\mathrm{g}}
= \frac{n_{0} e}{2 \pi \varepsilon B}
\ln \left( \frac{e^{- C}}{2ka^{\ast}} \right),
\label{velocity1}
\end{equation}
where $\varepsilon$ is the dielectric constant of GaAs and $C$ is 
the form factor of the order of 1, which depends on 
the electron density profile around the edge channel.
The $a^{\ast}$ is the characteristic width in which the EMP is 
mainly confined transversely to the edge channel.
On the other hand, in the case of screened 2DEG with a surface 
metal [Fig.~\ref{fig_screening}(a)], \cite{johnson:2003} 
$v_{\mathrm{g}}$ is drastically reduced due to the screening of
the in-plane electric field.
When the depth of the 2DEG below the metal, $d$, is much smaller
 than the width of the edge channel, $a$, $v_{\mathrm{g}}$ 
is given by
\begin{equation}
v_{\mathrm{g}} = \frac{n_{0} e}{\varepsilon B} \frac{d}{a}, 
\; \; \; d \ll a.
\label{velocity2}
\end{equation}
Therefore, for the high-screening case, the velocity 
$v_{\mathrm{g}}$ is reduced by a factor of $d/a$, which
represents the degree of screening.

Figure \ref{fig_screening}(b) schematically shows how the group 
velocity $v_{\mathrm{g}}$ depends on the screening
represented by the depth $d$.
In the limit of $d \ll a$, $v_{\mathrm{g}}$ is reduced in 
proportion to the degree of screening [Eq.~(\ref{velocity2})].
In the opposite limit of $d \gg a$, the surface metal doesn't 
affect the EMPs and $v_{\mathrm{g}}$ 
is expected to be independent of $d$ and approach the value 
given by Eq.~(\ref{velocity1}).
Although we do not know the analytical expression  for $d \sim a$, 
$v_{\mathrm{g}}$ is expected to vary 
as shown by the dashed line in Fig.~\ref{fig_screening}(b).
Since the surface of our device is partially covered with metallic 
gates, our experiment corresponds to 
the weak-screening regime ($d \sim a$).
Therefore, the velocity $v_{\mathrm{g}}$ is sensitive to the 
geometry of electrostatic boundary condition.

\begin{figure}[h]
\begin{center}
\includegraphics[width=77mm]{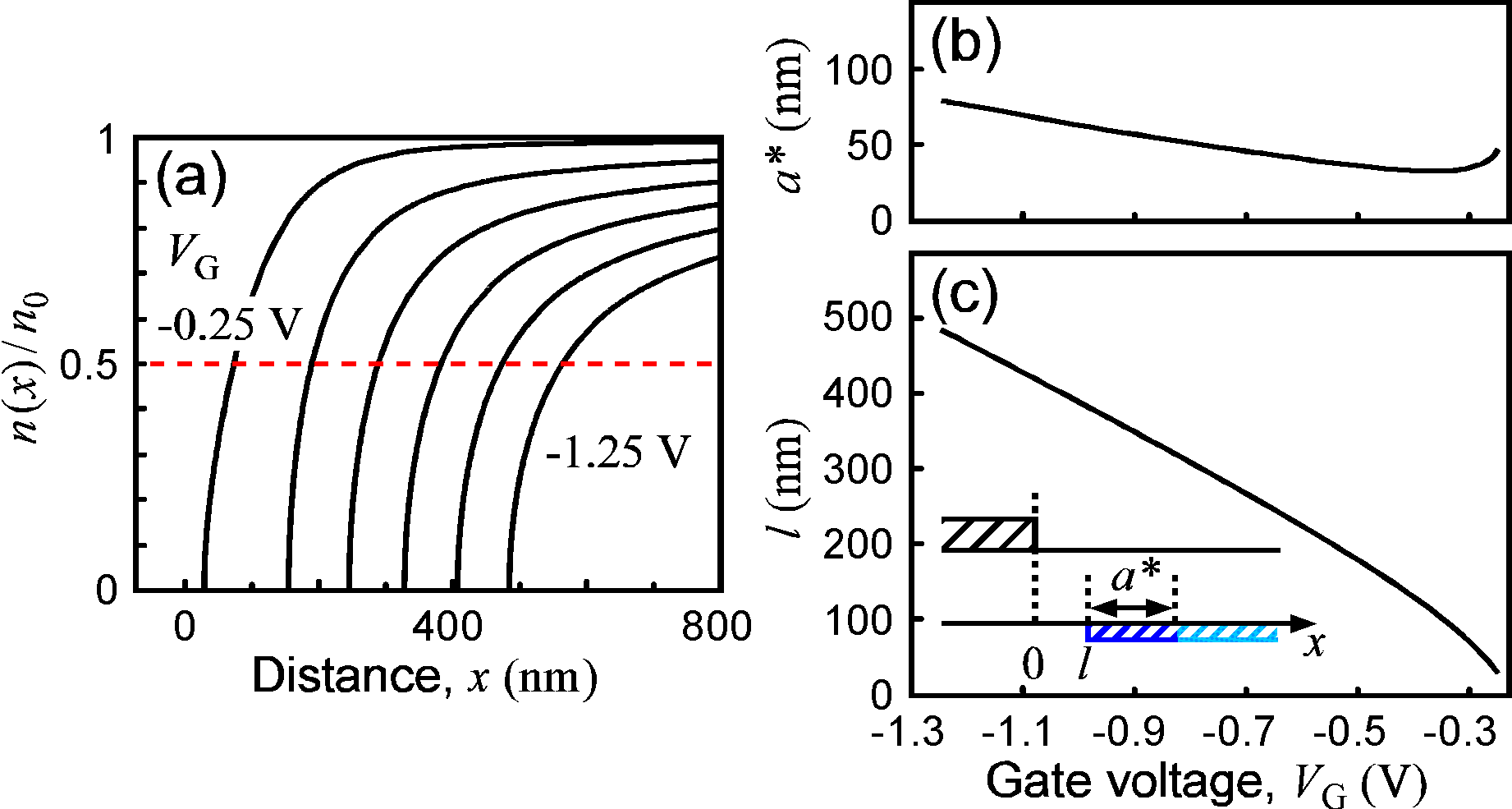}
\caption{(Color online) (a) Normalized electron density profiles
at the distance $x$ from the gate for various $V_{\mathrm{G}}$.
The pinch-off voltage of $-0.2$ V and the 2DEG depth of $110$ nm 
are used for the calculation.
(b) Characteristic width $a^{\ast}$ as a function of 
$V_{\mathrm{G}}$.
(c) Depletion length $l$ as a function of $V_{\mathrm{G}}$.
The inset shows a schematic cross-section around the edge 
channel.}
\label{fig_deplete}
\end{center}
\end{figure}

In order to evaluate the spatial profile of the edge channels 
in our device, we used the analytical formula given by Larkin 
and Davies for an edge channel induced at a depth $d$ by a 
semi-infinite metallic gate. \cite{ivan:1995}
Fig.~\ref{fig_deplete}(a) shows the normalized electron density 
profiles, $n(x)/n_{0}$, for various $V_{\mathrm{G}}$ 
at $d = 110$ nm at zero magnetic field, 
where $n_{0}$ is the bulk electron density and $x$ is the lateral 
distance from the edge of the gate.
We take for the characteristic width $a^{\ast}$ the distance 
between the onset of finite $n(x)$ and the half maximum, 
$n(x) = 0.5 \ n_{0}$, \cite{aleiner:1994} and for the depletion 
length, $l$, the distance between the gate edge and the position 
of the $n(x)$ onset.
They are  plotted as a function of $V_{\mathrm{G}}$ in 
Figs.~\ref{fig_deplete}(b) and (c).
For sufficiently negative $V_{\mathrm{G}}$ ($< -0.3$ V), both 
$a^{\ast}$ and $l$ increase monotonically 
with decreasing $V_{\mathrm{G}}$.
Here, we can find two opposite effects on the velocity 
$v_{\mathrm{g}}$.
On one hand, the width $a^{\ast}$ is related to the electrostatic 
edge potential profile around the edge channel. 
\cite{chklovskii:1992}
Eqs.~(\ref{velocity1}) and (\ref{velocity2}) suggest that as 
$a^{\ast}$ increases, the velocity $v_{\mathrm{g}}$ reduces.
On the other hand, the depletion length $l$, which represents 
the distance between the  edge channel and the metallic gate,
can be regarded as the measure of screening.
Analogous to the $d$-dependence of $v_{\mathrm{g}}$ shown in 
Fig.~\ref{fig_screening}(b), this implies that $v_{\mathrm{g}}$ 
should increase as $l$ increases.
The observed $V_{\mathrm{G}}$-dependence indicates that 
the variation in the depletion length $l$ (degree of screening) 
has a larger effect on 
the velocity $v_{\mathrm{g}}$ as compared to the change in the width 
$a^{\ast}$ (electrostatic edge potential profile).

Actually the two effects are intricately related to each other.
The edge geometry ($a^{\ast}$ and $l$) is determined by the electrostatic
 solution for the given boundary condition. \cite{chklovskii:1992}
Image charges induced in the gates by the EMPs affect the electrostatic 
edge geometry.
Therefore, the velocity $v_{\mathrm{g}}$ is determined self-consistently.
Although quantitative discussion requires further investigation, our 
results clearly demonstrate that the group velocity of the EMPs has been 
controlled electrically by changing the degree of screening.

\section{Summary}

We have successfully demonstrated voltage-controlled group velocity of
the EMPs in the quantum Hall regime at the bulk filling factor 
$\nu = 2$.
The group velocity strongly depends on the degree of screening caused 
by the metallic gate.
Although we have investigated the group velocity only for $\nu = 2$,
we expect similar variation of the velocity for other filling factors.

Our experimental technique will be useful for conducting electronic 
interferometric experiments in the pulse mode as well as for 
studying electron dynamics in edge channels.
For example, even if one of the (spin-resolved) Landau levels is
selectively excited, charges in an edge channel are strongly
affected by adjacent edge channels and the injected charge is 
expected to be fractionalized into various edge 
channels. \cite{berg:2009}
Such intriguing behavior may be clarifed by further investigating 
the local and time-resolved potential measurement developed in this
work.

\begin{acknowledgments}
We thank N. Kumada for fruitful discussions and M. Ueki for experimental
support.
This work was partially supported by the Strategic Information and 
Communications R\&D Promotion Programme (SCOPE) from the MIC of Japan, 
a Grant-in-Aid for Scientific Research (21000004) from the MEXT of Japan,
and the Global Center of Excellence Program from the MEXT of Japan through 
the ``Nanoscience and Quantum Physics" Project of the Tokyo Institute 
of Technology.
\end{acknowledgments}

\end{document}